\documentclass[reprint,superscriptaddress]{revtex4-2}
\usepackage{graphicx} %
\usepackage{amsmath}
\usepackage{physics}
\usepackage{xcolor}
\usepackage{amsfonts}
\usepackage[colorlinks]{hyperref}
\begin{document}

\title{Single snapshot non-Markovianity of Pauli channels}
\author{Alireza Seif}
\affiliation{IBM Quantum, IBM T.~J.~Watson Research Center, Yorktown Heights, NY 10598, USA}
\author{Moein Malekakhlagh}
\affiliation{IBM Quantum, IBM T.~J.~Watson Research Center, Yorktown Heights, NY 10598, USA}
\author{Swarnadeep Majumder}
\affiliation{IBM Quantum, IBM T.~J.~Watson Research Center, Yorktown Heights, NY 10598, USA}
\author{Luke~C.~G.~Govia}
\altaffiliation{Present address: CMC Microsystems, Waterloo, ON, Canada.}
\affiliation{IBM Quantum, IBM Almaden Research Center, San Jose, CA 95120, USA}

\begin{abstract}
Pauli channels are widely used to describe errors in quantum computers, particularly when noise is shaped via Pauli twirling. A common assumption is that such channels admit a Markovian generator, namely a Pauli-Lindblad model with non-negative rates, but the validity of this assumption has not been systematically examined. Here, using CP-indivisibility as our criterion for non-Markovianity, we study multi-qubit Pauli channels from a single snapshot of the dynamics. We find that while the generator always has the same structure as the standard Pauli-Lindblad model, the rates may be negative or complex. We show that random Pauli channels are almost always non-Markovian, with the probability of encountering a negative rate converging doubly exponentially to unity with the number of qubits. For physically motivated noise models shaped by Pauli twirling, including single-qubit over-rotations and two-qubit amplitude damping errors, we find that negative rates are generic, even when the underlying physical noise is Markovian. We generalize probabilistic error amplification and cancellation to non-Markovian generators, and quantify the sampling overhead introduced by negative and complex rates. Experiments on superconducting qubits confirm that allowing negative rates in the learned noise model yields more accurate predictions than restricting to non-negative rates.
\end{abstract}

\maketitle
\section{Introduction}
It is extremely challenging to completely isolate a quantum system. In general, the dynamics of an open quantum system, i.e., a system interacting with an unknown environment, are described by a linear, completely positive trace-preserving (CPTP) map, also known as a quantum channel~\cite{breuer2002theory}. However, in many-body systems, working directly with a quantum channel is often impractical. Instead, it is preferable to work with the generator of the channel, similar to how we typically work with the Schr\"odinger equation and the Hamiltonian, rather than the full unitary operator that describes the evolution of a closed system. When the interactions between the system and its environment are weak, and memory effects in the dynamics can be neglected, the generator evolution can be accurately captured by what is known as a Markovian master equation. Determining whether an observed evolution admits a Markovian generator therefore has both fundamental and practical significance, though the structure of such dynamics can be subtler than it first appears. 

There have been several proposals for characterizing Markovianity of the evolution~\cite{rivas2014quantum,RevModPhys.88.021002}, including criteria based on CP-divisibility, i.e., the divisibility of the evolution into intermediate physical CPTP maps~\cite{PhysRevLett.105.050403,PhysRevLett.101.150402,PhysRevA.89.042120}, and those based on information backflow \cite{PhysRevLett.103.210401,wissmann2015generalized}, information lost about the system's steady-state under the Markov approximation \cite{Keefe2025}, and multi-time process correlations~\cite{PhysRevLett.120.040405,PhysRevLett.123.040401}. These criteria form a hierarchy of increasingly strict notions of Markovianity~\cite{li2018concepts}. In general, these measures are not equivalent and their results must be interpreted with care~\cite{PhysRevA.83.052128}. Perhaps counterintuitively, even when individual channels are Markovian, their probabilistic mixtures can lead to dynamics without a Markovian generator~\cite{PhysRevLett.101.150402,PhysRevA.101.062304,PhysRevA.91.012104,megier2017eternal,PhysRevA.96.032111,breuer2018mixing}. That is, unlike convex combination of the generators which always leads to a CPTP map, a probabilistic mixture of Markovian channels can lead to an evolution without a Markovian generator. This problem has been widely studied for single qubits in the case of Pauli channels~\cite{JAGADISH2020126907,megier2017eternal}, where it has been shown that mixing single qubit Pauli channels can result in a non-Markovian (CP-indivisible) channel. This has been extended to qudits, where again a convex combination of Markovian channels has been shown to be non-Markovian~\cite{PhysRevA.91.012104,siudzinska2020quantum}. While these results provide interesting insights about the geometry of Markovian channels, the examples may appear artificial, and it has only been recently that such mixtures for qubits have been engineered in an experiment~\cite{PhysRevA.109.042419}.

In this work, we study the non-Markovianity of multi-qubit Pauli channels. Our focus on Pauli channels is motivated by their practical success in describing errors in real-world quantum computers when Pauli twirling is applied~\cite{PhysRevA.94.052325,PhysRevX.11.041039,van2023probabilistic,kim2023evidence,chen2025disambiguating,aharonov2025reliable}. We use the criteria of Refs.~\cite{PhysRevA.89.042120,PhysRevLett.101.150402} that attribute non-Markovianity to the presence of negative rates in the master equation, a sufficient condition for CP-indivisibility. We show that while not all Pauli channels admit a Markovian generator, commonly referred to as a Pauli-Lindblad generator~\cite{van2023probabilistic}, they can be described by what we term a \emph{Pauli pseudo-Lindblad} generator of the same form, but with negative or complex rates. We also show that most random Pauli channels are non-Markovian, with the probability of encountering a non-Markovian channel increasing doubly exponentially with the number of qubits, and exponentially with the process infidelity. 

From a practical perspective, we show that it is common to observe non-Markovian Pauli channels when Pauli twirling is used to shape natural error into Pauli channel form. That is, Pauli-twirled error channels arising from common sources of error in quantum computers do not typically have a Markovian generator. Based on this observation, we propose a modified noise learning and mitigation approach tailored to such non-Markovian errors. Finally, we implement this protocol experimentally with superconducting qubits and characterize the resulting non-Markovian Pauli channel. We demonstrate that accounting for this type of non-Markovianity increases the accuracy of error mitigation without introducing additional implementation complexity of the mitigation protocol.

\section{Setup}
A Pauli channel $\mathcal{E}$ describes a probabilistic mixture of Pauli operators
\begin{equation}\label{eq:paulichannel}
    \mathcal{E}(\rho) = \sum_k p_k P_k \rho P_k, 
\end{equation}
where $p_k$ are the Pauli error probabilities satisfying $\sum_k p_k =1$ and $P_k \in \{I,X,Y,Z\}^{\otimes n}$ is an $n$-qubit Pauli operator. 

The central question in this work is whether the evolution  $\mathcal{E}$ can originate from a Markovian generator. That is, whether there exist $\mathcal{L}$ such that $e^\mathcal{L} = \mathcal{E}$ where 
\begin{equation}
    \mathcal{L}(\rho) = \sum_{k} \lambda_k (L_k \rho L_k^\dagger - \frac{1}{2}\{L_k^\dagger L_k,\rho\}), \label{eqn:GKSL}
\end{equation}
is the Gorini-Kossakowski-Sudarshan-Lindblad (GKSL) generator. When the Lindblad rates are non-negative, i.e., $\lambda_k\geq0$, the evolution $e^{\mathcal{L}t}$ is guaranteed to be a completely-positive trace-preserving map. Note that by restricting our study to Pauli channels we can drop the Hamiltonian evolution term of the full GKSL generator and focus solely on the dissipative evolution described by Eq.~\eqref{eqn:GKSL}.

This question is naturally connected to a widely used mathematical characterization of Markovianity known as completely positive (CP) divisibility~\cite{wolf2008dividing,PhysRevLett.101.150402}. A dynamical map $\Lambda(t,0)$ is CP-divisible if it can be written as
\begin{equation}
    \Lambda(t_2,0) = \Lambda(t_2,t_1)\,\Lambda(t_1,0),
    \quad \forall \; t_2 \ge t_1 \ge 0,
\end{equation}
where each intermediate map $\Lambda(t_2,t_1)$ is completely positive (CP). Loss of CP-divisibility is often taken as a signature of non-Markovianity. While other definitions such as those based on information backflow~\cite{PhysRevLett.103.210401} or distinguishability measures are also used, CP-divisibility remains one of the most operationally and mathematically tractable frameworks. Therefore,  in this work, we refer to channels $\mathcal{E}$ that fail to admit a generator $\mathcal{L}$ such that all $\lambda_k$ are non-negative as non-Markovian channels.

As Pauli channels~\eqref{eq:paulichannel} are diagonal in the Pauli basis, their generators admit a simple form. As we show in the following, the generator of a Pauli channel is always of the form \begin{equation}\label{eq:pauli-lindblad}
    \mathcal{L}(\rho) = \sum_k \lambda_k P_k \rho P_k, 
\end{equation}
where $P_0=I$ indicates the identity component. The trace-preserving property then implies that $\lambda_0 = -\sum_{k\geq 1} \lambda_k$ and therefore $\mathcal{L}$ can be expressed as
\begin{equation}     \label{eq:pauli-lindblad1}
    \mathcal{L}(\rho) = \sum_{k\geq1} \lambda_{k} (P_k \rho P_k -\rho).
\end{equation}
This generator is of the form of a GKSL generator with $L_k = P_k$, where the $P_k$ are Pauli and therefore self-inverse. If it is Markovian with $\lambda_{k}\geq0$, then it is a Pauli-Lindblad generator as studied in Ref.~\cite{van2023probabilistic}. Consequently, to assess the Markovianity of Pauli channels, we simply need to find $\lambda_k$ and assess if they are all non-negative. When the rates are not constrained to be non-negative, we refer to Eq.~\eqref{eq:pauli-lindblad1} as a Pauli pseudo-Lindblad generator, following the terminology introduced in Ref.~\cite{groszkowski2023simple}. Beyond its role in characterizing non-Markovianity, this generator also provides a convenient representation for noise modeling and error mitigation, as we discuss in Sec.~\ref{sec:applicatons}.

To find the generator, we need to evaluate the matrix logarithm of the channel, which simplifies to a standard logarithm in the Pauli transfer matrix (PTM) picture. In this picture, we have 
\begin{equation}
    \mathcal{E}(\rho) = \sum_i f_k P_k \Tr(P_k \rho),  
\label{eqn:E_from_fk}
\end{equation}
where $f_j = \sum h_{jk} p_k$, and $h_{jk} = 1$ if $[P_k,P_j]=0$ and $h_{jk} = -1$ otherwise. The relationship between the Pauli fidelities (or eigenvalues) $f_k$ and Pauli error probabilities $p_k$ can be expressed concisely in vector form as $\vec{f} = H \vec{p}$, where $H$ is the Walsh-Hadamard transform. Since Pauli channels have diagonal PTMs, we can find their generator by taking the logarithm of the fidelities and find 
\begin{equation}
    \mathcal{L}(\rho) = \sum_k \log(f_k) P_k \Tr(P_k \rho).
\label{eqn:L_from_fk}
\end{equation}
We then express this equation in the Pauli pseudo-Lindblad form~\eqref{eq:pauli-lindblad1} by applying the inverse Walsh-Hadamard transform to find $\lambda_k = \frac{1}{D} \sum_{j} h_{kj} \log(f_j)$ and $D=4^n$. Therefore, we can concisely express the rates in vector form as  
\begin{equation}
    \vec{\lambda} = \frac{1}{D} H \log(H \vec{p}).
\label{eq:lambda-log-fid}
\end{equation}
Note that the Pauli-Lindblad equation \eqref{eq:pauli-lindblad1} is guaranteed to continuously generate a CPTP map at all times if the rates $\lambda_k$ are non-negative. However, we are only interested in the channel $e^{t\mathcal{L}}$ at $t=1$ (or its integer multiples). Therefore, having non-negative rates is only sufficient, but not necessary, to obtain a CPTP map from $e^\mathcal{L}$. 
In fact, as we show, for general Pauli channels, the rates can be negative or even complex and still generate a valid channel at $t=1$. Therefore, in the most general setting even non-Hermitian Pauli pseudo-Lindblad generators are allowed. 

We illustrate this point by several examples, starting from a single-qubit Pauli channel, going to random multi-qubit Pauli channels, and finally to physically motivated examples of error channels affecting quantum operations. The Pauli channels in the final examples are more operationally useful tools for characterizing and mitigating error, and as we will discuss that the non-Markovianity there has less to do with the underlying physical process and more with how tools such as randomized compiling are used to shape the errors~\cite{PhysRevA.94.052325,PhysRevX.11.041039}.

\section{Case studies}
Pauli channels are commonly used in the discussion of quantum errors. When the nontrivial error probabilities $p_k$ are small, the corresponding Pauli fidelities  $f_k$ are close to unity, and the Lindblad rates $\lambda_k$ are therefore real. For larger error rates, however, the fidelities $f_k$ can become negative, leading to complex Lindblad rates. We begin by considering an example in which the Lindblad rates may become complex, and then restrict our attention to the low-error regime, where all $\lambda_k$ are real.
\begin{figure}
    \centering
    \includegraphics[width=\columnwidth]{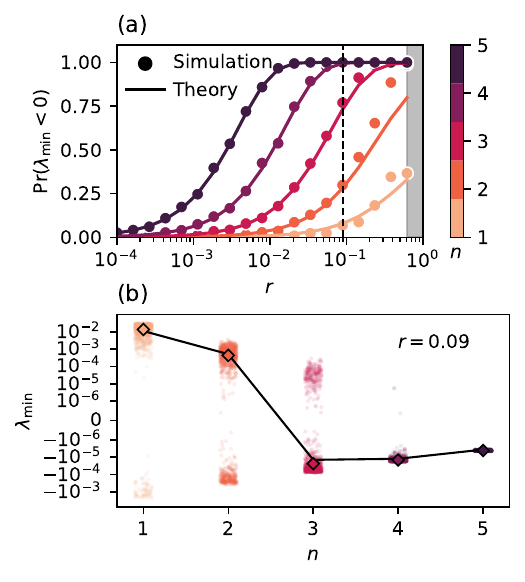}
    \caption{ Non-Markovianity  of random Pauli channels. (a) The probability of having a negative Lindblad rate $p(\lambda_{\min}<0)$ as a function of infidelity $r$ grows with increasing system size $n$. Numerical simulations (circles) agree well with our analytical results in Eq.~\eqref{eq:problambdamin}. In the shaded grey region the rates for the smallest system size become complex and are not shown. (b) The minimum Lindblad rate $\lambda_{\min}$ at $r=0.09$ (dashed line in the top panel). The scatter points are samples from numerical simulations and the diamond markers indicate their mean values. Our analytical result for the mean in Eq.~\eqref{eq:explambdamin} agrees well with the numerical results. For larger $n$ the values concentrate around their mean. While negative values become more common at larger $n$, their magnitude gets smaller.
    \label{fig:rand-channel}}
\end{figure}
\subsection{A simple single-qubit example}
Consider the single-qubit Pauli channel family 
\begin{equation}
    \mathcal{E}(\rho) = (1-p_x-p_y-p_z) \rho + p_x X\rho X + p_y Y \rho Y + p_z Z \rho Z, 
\end{equation}
and choose $(p_x,p_y,p_z)=(0,1-2p,p)$ with $p\in(0,\frac{1}{2})$. For this parameter choice, the no-error probability $p$ is strictly smaller than the total error probability $1-p$, and the channel applies an error more often than it preserves the state. By performing the Walsh-Hadamard transform we find that $(f_x,f_y,f_z)=(2p-1,1-2p,4p-1)$.  Therefore, there is always at least one negative Pauli fidelity, which results in complex rates as $\lambda_k$ are linear combinations of complex-valued $\log(2p-1)$, $\log(1-2p)$, and $\log(4p-1)$. 

Given that deterministic Pauli operations ($\mathcal{E}(\rho) = P\rho P$) also have negative $f_k$, one may wonder if in large-error cases a single-qubit Pauli channel with negative fidelities can always be decomposed as a composition of a deterministic unitary operation and a channel with positive fidelities.  For $p<\frac{1}{4}$, we can indeed remove the negative fidelities, and consequently the complex rates, by decomposing the channel into two channels, one of which is unitary. Specifically, for our case the evolution can be decomposed into applying a channel with $(p_x,p_y,p_z)=(p,p,0)$, followed by a deterministic application of a $Y$ unitary. However, for $p>\frac{1}{4}$ this simple decomposition no longer works and the evolution is not equivalent to a Pauli channel with positive fidelities followed by a unitary Pauli operation. Thus, complex generator rates are a more fundamental property of Pauli channels than simply being a ``hidden'' unitary error that can be factored out.

\subsection{Random Pauli Channels}
\label{sec:randompauli} We next consider random Pauli error channels 
\begin{equation}
    \mathcal{E}(\rho) = (1-r)\rho + \sum_k p_k P_k \rho P_k, 
\end{equation}
where the coefficients $p_k$ are drawn uniformly at random subject to $\sum_{k\geq 1} p_k = r$ and $0 \leq p_k \leq 1$. For convenience, we define $p_0 = 1-r$, where the parameter $r$ is the infidelity of the channel. Throughout this example we focus on the low-error regime $r \ll 1$, in which all Pauli fidelities $f_k$ remain positive.

To decide the Markovianity of such a channel, we must verify that all Lindblad rates satisfy $\lambda_k > 0$ for $k \geq 1$. As shown in Eq.~\eqref{eq:lambda-log-fid}, the rates $\lambda_k$ are linear functions of $\log(f_k)$, and fidelities $f_k$ themselves depend linearly on the error probabilities $p_k$. This simple structure allows us to study, both analytically and numerically, the distribution of Lindblad rates induced by random choices of $p_k$, and thereby characterize the statistics of non-Markovianity in these channels.

As shown in Fig.~\ref{fig:rand-channel}, the probability of having a non-Markovian channel grows rapidly with system size.  This probability is given by the probability that the minimum Lindblad rate $\lambda_{\min} = \min_{k>0} \lambda_k$ is negative (see Fig.~\ref{fig:rand-channel}a). In Appendix~\ref{app:random}, we show that this probability converges doubly exponentially to unity with qubit number, as \begin{equation}
\label{eq:problambdamin}
    {\rm{Pr}}(\lambda_{\rm{min}}<0) \approx 1-\exp(-\frac{D r}{4(1+r)}).
\end{equation}
We also compute the average value of the minimum rate
 \begin{equation}\label{eq:explambdamin}
     \expval{\lambda_{\min}} = \frac{r}{2 D^2-1} \left(-\left(\left(D-4\right) r\right)-4 e^{-D} (r+1)+4\right).
\end{equation}
and find that, at fixed $r$, it becomes increasingly concentrated around negative values as the number of qubits $n$ increases (see Fig.~\ref{fig:rand-channel}b). Together, these results show that non-Markovianity becomes typical for random Pauli channels as the system size increases, even in the low-error regime. While the probability of a negative Lindblad rate rapidly approaches unity, the magnitude of the most negative rate decreases and concentrates around a small negative value. Thus, in large systems, non-Markovianity is generic but typically weak.

\subsection{Twirled physical noise channels}
\label{subsec:twirled_phys}
Here we consider three physically motivated examples of Pauli noise that may arise in experiments. All examples concern Clifford gates that are Pauli twirled, which are relevant for applications in quantum benchmarking, error mitigation, and error correction. In these examples, we denote the ideal gate channel by $\mathcal{G}$ and its  noisy implementation by $\tilde{\mathcal{G}}$. Without loss of generality, we define the Pauli-twirled noise channel $\mathcal{E}$ and assume it acts before the gate, i.e., $\tilde{\mathcal{G}}=\mathcal{G}\circ \mathcal{E}$, which implies that
\begin{equation}\label{eq:noisegate}
    \mathcal{E} = \mathcal{G}^{-1}\circ\tilde{\mathcal{G}}. 
\end{equation}
We emphasize that the non-Markovianity of $\mathcal{E}$ in these examples does not reflect genuine memory effects in the evolution~\cite{megier2017eternal}. Instead, it is an artifact of separating the noise from the ideal evolution and applying Pauli twirling. While this notion of non-Markovianity in such cases does not provide physical insight, we show in Sec.~\ref{sec:applicatons} that it can be useful for more accurate noise modeling and error mitigation in quantum circuits.
\begin{figure}
    \centering
    \includegraphics[width=\columnwidth]{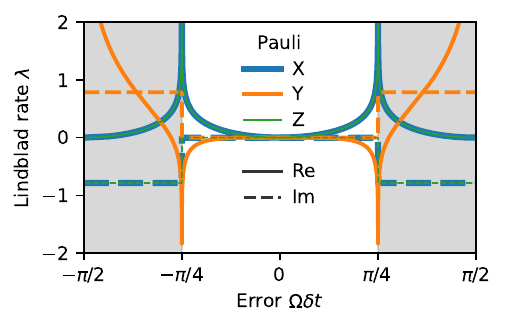}
    \caption{Lindblad rates for a noisy Hadamard gate with timing error. Error in timing $\delta t$ results in over/under rotations. For $\abs{\Omega\delta t}<\pi/4$, where $\Omega$ is the strength of the Hamiltonian evolution, one of the rates $\lambda_Y$ is never positive. For larger errors  $\pi/4<\abs{\Omega\delta t}<\pi/2$, all of the rates are complex (shaded region). }
    \label{fig:had-error}
\end{figure}
\subsubsection{Hadamard gate with over rotation}
The first example we consider is a Hadamard gate subject to over- and under-rotation errors. The Hadamard gate is generated by the Hamiltonian
\begin{equation}
    H_{1} = \frac{\Omega}{\sqrt{2}} (X + Z).
\end{equation}
The unitary evolution $U = \exp(-i H_1 t)$ implements the ideal Hadamard operation when $\Omega t = \pi/2$. However, if the gate duration is imperfect, $t \to t + \delta t$, coherent rotation errors are introduced. Pauli twirling the faulty gate and extracting the corresponding error channel yields
\begin{equation}
    \mathcal{E}_1 (\rho) = (1 - 2p)\rho + p X \rho X + p Z \rho Z,
\end{equation}
where $p = \frac{1}{2}\sin^2(\Omega \delta t)$. The Markovianity of this Pauli channel, considered independently of the underlying gate error and Pauli-twirling procedure, was previously studied in Ref.~\cite{PhysRevA.101.062304}. 
Such a channel can be expressed as $\mathcal{E}_1  = \exp(\mathcal{L}_1)$ where 
\begin{equation}
    \mathcal{L}_1(\rho) = \sum_{k \in \{X,Y,Z\}} \lambda_k (P_k \rho P_k -\rho), 
\end{equation}
where $\lambda_X=\lambda_Z=-\frac{1}{4}\log(1-4p)$, and $\lambda_Y = -\frac{1}{2}\log(1-2p)+\frac{1}{4}\log(1-4p)$. For $p>\frac{1}{4}$ these rates are complex. However, for $0<p<\frac{1}{4}$, $\lambda_Y$ is always negative, see Fig.~\ref{fig:had-error}. Notably, this negative rate persists even in the low-error regime with $p\ll 1$. 

Intuitively, as elucidated by the Magnus expansion, a continuous time process with $X$ and $Z$ dissipators necessarily generates a $Y$ error. However, the Pauli-twirled error channel $\mathcal{E}_1$ contains no $Y$ Pauli component. As a result, a negative $Y$ rate is required so that $\exp(\mathcal{L}_1)$ reproduces the channel exactly. Of course, although $\mathcal{L}_1$ contains a negative rate, the map $\exp(s\mathcal{L}_1)$ is completely positive and trace preserving at $s=1$ as required. However for $0 < s < 1$ this is no longer the case.

\subsubsection{$X_{\pi}$ gate with amplitude and phase error}
We consider the implementation of an $X_{\pi}$ gate with coherent amplitude and phase errors. The Hamiltonian is given by 
\begin{equation}
    H_X = (\Omega+\delta \Omega) X + \epsilon Z, 
\end{equation}
where $\Omega$ is the ideal drive amplitude, $\delta \Omega$ is the amplitude error, and $\epsilon$ characterizes the off-axis error originating from phase errors in the drive. 

\begin{figure}
    \centering
    \includegraphics[width=\columnwidth]{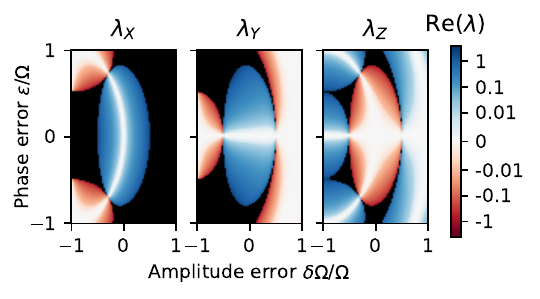}
    \caption{Lindblad rates for a noisy $X_{\pi}$ gate with coherent amplitude and phase error. Phase errors can lead to off-axis errors $\epsilon$, whereas deviations from ideal amplitude $\Omega$ by $\delta \Omega$ cause over/under rotations errors. In the small noise regime, all the rates are real, with $\lambda_Z$ being negative (red region). For larger errors some of the rates become complex (black region). }
    \label{fig:x-gate}
\end{figure}

Ideally, when $\delta=\epsilon=0$, at $\Omega t = \pi/2$ this evolution realizes a Pauli $X$ gate. In the presence of amplitude and phase errors, however, the realized evolution deviates from the ideal operation. As before, we isolate the noise channel according to Eq.~\eqref{eq:noisegate} and then apply Pauli twirling to obtain the corresponding Pauli error channel. Depending on the values of $\epsilon$ and $\delta \Omega$, the resulting twirled error channel exhibits complex or negative Lindblad rates as shown in Fig.~\ref{fig:x-gate}. In the low error regime, we observe that $\lambda_X$ and $\lambda_Y$ remain positive, while $\lambda_Z$ becomes negative. For larger errors some of the rates become complex. As the expressions in this case are rather complicated we refer the reader to Appendix~\ref{appendix:twirled_phys} for details. 

\subsubsection{$ZZ_{\pi/2}$ gate with $T_1$ damping}
\begin{figure}
    \centering
    \includegraphics[width=1\columnwidth]{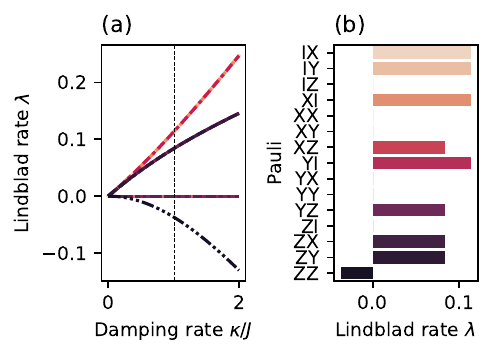}
    \caption{Lindblad rates for the two-qubit $ZZ_{\pi/2}(\pi/4)$ gate with amplitude damping error. (a) Lindblad rates (colors corresponding to the Pauli labels in the right panel) as a function of the damping rate $\kappa/J$, where $J$, the strength of the gate, sets the time scale. For all values of $\kappa/J$, one of the rates $\lambda_{ZZ}$ is nonpositive, and its magnitude grows with increasing $\kappa/J$. (b) Comparing different Lindblad rates  at $\kappa/J=1$ corresponding to the dashed line in panel (a).   
   }
    \label{fig:rzz-gate}
\end{figure}

The next example we consider is the $ZZ_{\pi/2}$ gate with amplitude damping error. Prior to Pauli twirling, the evolution is governed by
\begin{equation}
    \partial_t\rho = - iJ [Z_1 Z_2,\rho] + \sum_{j=1,2} \kappa (\sigma^-_j \rho  \sigma^+_j - \frac{1}{2} \{\sigma^+_j\sigma^-_j, \rho\}),
\end{equation}
where $J$ is the coupling strength and $\kappa$ is the amplitude damping rate, which for simplicity is taken to be identical for both qubits. For $Jt = \pi/4$, this evolution generates a Clifford gate that is equivalent to a CZ gate up to local rotations. After Pauli twirling, we find that all Pauli fidelities remain positive. However, one of the Lindblad rates, $\lambda_{ZZ}$, is always negative, as shown in Fig.~\ref{fig:rzz-gate}. By expanding the Lindblad rates to second order in $\kappa/J$, we find $\lambda_{ZZ} = -\frac{ \pi^2 \kappa^2}{256 J^2}$~\cite{malekakhlagh2025efficient}. 
Notably, although the physical dynamics is generated by a Markovian Lindbladian, the corresponding Pauli-twirled channel exhibits negative Lindblad rates. Moreover, the error process in this example is purely dissipative in nature, demonstrating that negative rates can occur even without a coherent (unitary-generating) error source.
\section{Applications}\label{sec:applicatons}
In this section, we examine the implications of non-Markovianity for applications of Pauli noise models in simulation and quantum error mitigation. We recall that Pauli channels are probabilistic mixtures of Pauli operations. Consequently, implementing such channels experimentally requires sampling Pauli operators according to their associated probabilities. However, sampling from an exponentially large, unstructured distribution is computationally difficult.

This motivates the use of a generator, and in particular a sparse generator, i.e., one with at most a polynomial number of terms. In practice, it is well motivated to assume such sparsity, since the channels encountered in realistic settings typically originate from local physical interactions~\cite{van2023probabilistic}. In the Pauli pseudo-Lindblad model~\eqref{eq:pauli-lindblad1}, all generator terms commute, which implies that the channel factorizes as
\begin{equation}
    \mathcal{E}(\rho) = \prod_k \mathcal{E}_k(\rho)
    = \prod_k e^{\mathcal{L}_k}(\rho),
\end{equation}
where
\begin{equation}
    \mathcal{L}_k(\rho) = \lambda_k \left(P_k \rho P_k - \rho\right).
\end{equation}
As a result, the channel $\mathcal{E}$ can be implemented by sequentially applying the individual maps $\mathcal{E}_k$. We emphasize again that the sparsity assumption in the generator is more natural as physical locality constraints manifest as sparse generators but not necessarily sparse channels due to the exponential map that connects generators and channels.

The cost of implementing $\mathcal{E}$ is therefore determined by how efficiently each factor $\mathcal{E}_k$ can be realized. When all rates $\lambda_k$ are real and positive, each $\mathcal{E}_k$ corresponds to a physical (Markovian) Pauli channel and admits a straightforward probabilistic implementation. However, negative or complex rates introduce a sign problem, which requires a quasi-probabilistic implementation and therefore incurring a sampling overhead.

We emphasize that even when the overall channel $\mathcal{E}$ is physical and described by valid probabilities, the individual factors $\mathcal{E}_k$ in this decomposition need not correspond to physical channels. This is the price for obtaining a sparse and efficiently implementable representation of the noise. We also note that the inverse channel $\mathcal{E}^{-1} = e^{-\mathcal{L}}$ has the same factorized form, with all rates $\lambda_k$ replaced by $-\lambda_k$. Consequently, any quasi-probabilistic procedure that implements $\mathcal{E}$ can be directly adapted to implement $\mathcal{E}^{-1}$. This observation underlies the principle of probabilistic error cancellation (PEC) with Pauli-Lindblad noise~\cite{van2023probabilistic}.

Next, we describe a (quasi-)probabilistic implementation of the individual terms
\begin{equation}
    \mathcal{E}_k(\rho) = w_k \rho + (1 - w_k) P_k \rho P_k,
\end{equation}
where $w_k = \frac{1}{2}(1 + e^{-2\lambda_k})$. To implement this channel, we consider three cases: (i) $\lambda_k > 0$, (ii) $\lambda_k < 0$, and (iii) $\lambda_k \in \mathbb{C}$. 

\paragraph{Case (i): $\lambda_k > 0$.}
This is the simplest case. When $\lambda_k > 0$, we have $w_k<1$ and  $1-w_k>0$. Therefore, $\mathcal{E}_k$ is a physical (Markovian) Pauli channel. It can be implemented by sampling and applying $I$ or $P_k$ with probabilities $w_k$ and $1 - w_k$, respectively. There is no sampling overhead in this case.

\paragraph{Case (ii): $\lambda_k < 0$.}
When $\lambda_k < 0$, we have $w_k > 1$ and $1 - w_k < 0$. The negative coefficient introduces a sign problem, and  we have to use a quasi-probabilistic approach. We rewrite the channel as
\begin{equation}
    \mathcal{E}_k(\rho) = \gamma_k \left[q_k \rho - (1 - q_k) P_k \rho P_k\right],
\end{equation}
where $\gamma_k = 2 w_k - 1$ and $q_k = \frac{w_k}{2 w_k - 1}$. The map is implemented by probabilistically applying $I$ or $P_k$ with probabilities $q_k$ and $1 - q_k$, respectively. In post-processing, samples in which $P_k$ was applied are multiplied by $-1$, and all expectation values are rescaled by $\gamma_k$. This rescaling increases the variance by a factor of $\gamma_k^2$, which is characteristic of the sign problem.

\paragraph{Case (iii): $\lambda_k \in \mathbb{C}$.}
This case is similar to case (ii), with additional modifications.
When $\lambda_k$ is complex, $w_k$ is also complex. We express the channel as
\begin{equation}\label{eq:gen_channel_k}
    \mathcal{E}_k (\rho) =  \gamma_k \left[q_k \rho + e^{i\phi_k} (1-q_k) P_k \rho P_k\right), 
\end{equation}
where $\gamma_k = \frac{w_k}{\abs{w_k}}(\abs{w_k} + \abs{1-w_k})$, $q_k = \abs{w_k}/(\abs{w_k} + \abs{1-w_k})$, and $\phi_k = \arg(\frac{1-w_k}{w_k})$. The prefactor $\gamma_k$ can be further simplified to
\begin{equation}\label{eq:simp_gamma}
    \gamma_k= e^{-\Re(\lambda_k )} (\abs{ \sinh (\lambda_k )} + \abs{\cosh (\lambda_k )} ).
\end{equation}

To implement this map, we again sample and apply $I$ or $P_k$ with probabilities $q_k$ and $1 - q_k$, respectively. In post-processing, samples in which $P_k$ was applied are multiplied by the phase $e^{i \phi_k}$, and all outcomes are rescaled by $\gamma_k$. In this case, $\gamma_k$ is generally complex, and the resulting variance overhead is given by $|\gamma_k|^2$. In this formulation, cases (i) and (ii) arise as special cases of case (iii). Moreover, when $\Im(\lambda_k) = m \pi / 2$ with $m \in \mathbb{Z}$, the map can be factorized into a composition of a unitary channel and a real-valued Pauli pseudo-Lindblad map, which simplifies the implementation.

Finally, to implement the full channel $\mathcal{E}(\rho) = \prod_k \mathcal{E}_k(\rho)$, we apply the above procedure independently for each $k$ and combine the results in post-processing by multiplying by the factor $\prod_k s_k \gamma_k$, where $s_k$ denotes the accumulated phase. The total overhead of the protocol is therefore determined by $\gamma = \prod_k \abs{\gamma_k}$. While directly implementing such noise models is useful for simulating open quantum systems (see e.g., ~\cite{PRXQuantum.4.040329,swain2025noise}), implementing amplified or inverse versions of these channels enables a wider range of applications in quantum error mitigation~\cite{PhysRevLett.119.180509,PhysRevX.7.021050,PhysRevX.8.031027,RevModPhys.95.045005}. In the following, we discuss two such applications in probabilistic error amplification (PEA) and PEC.

\subsection{Generalized PEA with Pauli pseudo-Lindblad models}
In probabilistic error amplification (PEA), one amplifies a learned Pauli noise channel by different factors and measures the expectation value of an observable of interest. By fitting the measured expectation values as a function of the amplification factor and extrapolating to the zero-noise limit, a procedure known as zero-noise extrapolation~\cite{PhysRevLett.119.180509,PhysRevX.7.021050}, one can estimate the noise-free expectation value~\cite{kim2023evidence}. 

The main building block of PEA is the ability to implement a desired Pauli noise channel. First, the noise channel is learned and represented using the sparse Pauli pseudo-Lindblad model. The Lindblad rates $\lambda_k$ are then scaled to $\alpha \lambda_k$, and the amplified noise is injected using the sampling scheme described above with the modified rates $\alpha \lambda_k$. Assuming a low-noise regime in which all $\lambda_k \in \mathbb{R}$, the overhead factor simplifies to
\begin{align}
    \gamma_{\rm{\alpha PEA}} =
     \exp(-2\alpha \sum_{\lambda_k<0}  \Re(\lambda_k))
\end{align}

Notably, amplifying a purely Markovian noise model with non-negative rates does not introduce a sampling overhead, whereas negative rates lead to an exponential increase in variance. As a result, for a fixed measurement budget, one can consider a bias-variance trade-off in which a Markovian approximation of the noise channel (e.g., by retaining only the positive-rate contributions) is amplified to reduce sampling overhead at the cost of introducing a controlled bias, the magnitude of which can be upper bounded \cite{Govia2025}.

\subsection{Generalized PEC with Pauli pseudo-Lindblad models}

In probabilistic error cancellation we are interested in undoing a known Pauli channel~\cite{PhysRevLett.119.180509,PhysRevX.8.031027}. Early formulations of PEC for Pauli--Lindblad noise were restricted to the case of positive rates $\lambda_k$. In that setting, the inverse generator $-\mathcal{L}$ contains negative rates, and the implementation of $e^{-\mathcal{L}}$ is analogous to case (ii) discussed above. Here, we consider the general case without imposing any restriction on the signs of $\lambda_k$. In this application, implementing the inverse channel corresponds to replacing each rate $\lambda_k$ with $-\lambda_k$.

It is then straightforward to apply our prescription for implementing a general Pauli channel $\mathcal{E}$ to the implementation of its nonphysical inverse $\mathcal{E}^{-1}$. Specifically, we use the general expressions in Eqs.~\eqref{eq:gen_channel_k} and \eqref{eq:simp_gamma}, with the substitution $\lambda_k \mapsto -\lambda_k$. Assuming a low-noise regime in which all $\lambda_k \in \mathbb{R}$, the total overhead for PEC simplifies to
\begin{align}
    \gamma_{\rm{PEC}} =
     \exp(2 \sum_{\lambda_k>0}  \Re(\lambda_k)).
\end{align}
 Notably, rates that are negative in the original channel do not contribute to the sampling overhead in PEC, since they become positive under inversion and admit a purely probabilistic implementation. By contrast, rates that are originally positive or complex lead to an exponential overhead. The interplay between non-Markovianity and quantum error mitigation overhead has also been explored in Refs.~\cite{PhysRevA.103.012611,endo2025non,nonmarkovmit2025wang}. 
 \begin{figure}
    \centering
    \includegraphics[width=\linewidth]{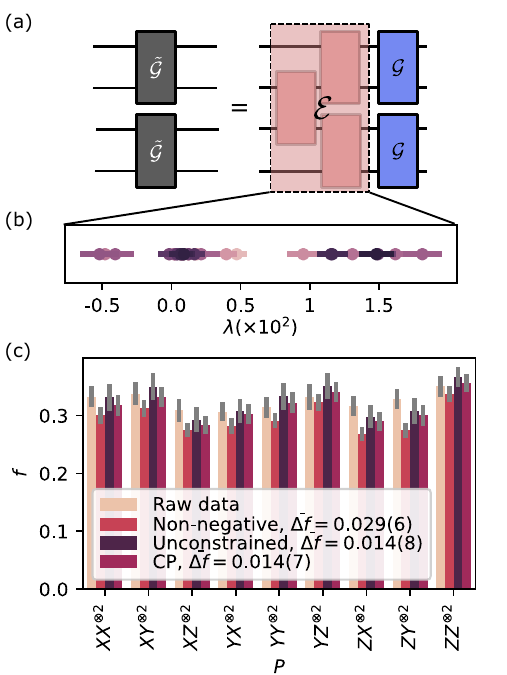}
    \caption{Four-qubit experiment on \emph{ibm\_pinguino1}. 
(a) The structure of each layer with two parallel noisy CNOT gates $\tilde{\mathcal{G}}$, and the equivalent decomposition into a quasi-local noise channel $\mathcal{E}$ acting on the ideal gates $\mathcal{G}$. (b) The inferred Lindblad rates $(\lambda)$, including statistically significant negative values. 
(c) Comparison between experimentally measured expectation values ($f$) of nonlocal Pauli observables for 8 repetitions of the layer and model predictions for different noise models. Numbers in the legend indicate the average error between the model and raw data.}
    \label{fig:chain-4}
\end{figure}
\section{Experimental results}
We demonstrate the practical relevance of Pauli pseudo-Lindblad generators through two experimental case studies, performed on systems of 4 and 12 qubits on \emph{ibm\_pinguino1}, respectively. In both experiments, we learn a local Pauli pseudo-Lindblad generator by estimating the Lindblad rates $\lambda_k$ associated with local Pauli operators, using measurements of local Pauli expectation values. We follow the procedure of Refs.~\cite{van2023probabilistic,van2024techniques}, and further details can be found in Appendix~\ref{app:learning}. 
To validate the learned models, we use high-weight observables obtained through a different marginalization of the measurement data than that used for learning. Specifically, rather than measuring all Pauli observables independently, we measure higher-weight Pauli strings and use their marginals to both train and test the noise model. For example, measurements in the all-X basis allow us to simultaneously estimate all expectation values of $P\in\{I,X\}^{\otimes4}$. We use the local ones such as $XXII$ in learning, while the non-local $XXXX$ is reserved for validation. This approach allows us to test the predictions of the learned model on observables that were not used directly in the learning stage, but for which the underlying data is the same, removing the impact of noise drift between learning and validation \cite{kim2025error}.

From the same experimental data, we infer three different noise models. The first is a Markovian Pauli-Lindblad model in which all Lindblad rates are constrained to be non-negative similar to Ref.~\cite{van2023probabilistic}. The second is an unconstrained Pauli pseudo-Lindblad model that allows negative rates and can therefore capture effective non-Markovian features of the noise. The third model also allows negative rates but explicitly enforces complete positivity (CP) of the resulting channel. While relaxing the nonnegativity constraint allows the model to be more expressive, it no longer guarantees that the generated channel is CPTP. Enforcing CP requires explicit construction of the full $n$-qubit channel, which scales exponentially with system size. For this reason, the CP model is considered only in the four-qubit experiment, where such a construction remains tractable.

For the four-qubit experiment, we consider a circuit consisting of layers of CNOT gates as shown in Fig.~\ref{fig:chain-4}a. We find that the learned unconstrained Pauli pseudo-Lindblad noise model exhibits statistically significant negative rates, as shown in Fig.~\ref{fig:chain-4}b.

We then compare the predictions of all three noise models against experimentally measured expectation values of nonlocal Pauli observables. As shown in Fig.~\ref{fig:chain-4}c, the unconstrained Pauli pseudo-Lindblad model provides improved agreement with the experimental data compared to the non-negative-rate model. Enforcing the CP constraint does not significantly alter the predictions, indicating that the improvement is mainly due to allowing negative rates and capturing non-Markovian features of the error channel.

\begin{figure}
    \centering
    \includegraphics[width=\linewidth]{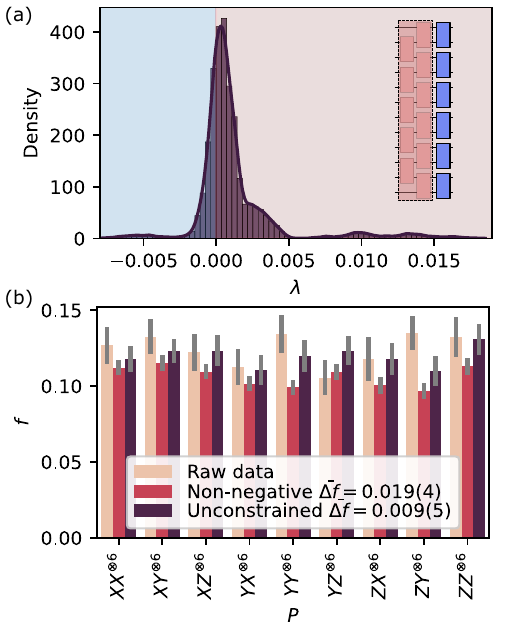}
    \caption{Twelve-qubit experiment on \emph{ibm\_pinguino1}. (a) Distribution of the inferred Lindblad rates $(\lambda)$ for the learned Pauli pseudo-Lindblad generator, showing both small and more pronounced negative values. The inset shows the circuit layer with six parallel  CNOT gates and the corresponding quasi-local noise model. 
    (b) Comparison between experimentally measured expectation values ($f$) of nonlocal Pauli observables for 8 repetitions of the layer and model predictions for different noise models. Numbers in the legend indicate the average error between the model and the raw data. We again observe that the unconstrained model provides more accurate predictions.}
    \label{fig:chain-12}
\end{figure}

For the twelve-qubit experiment, we consider layers of 6 parallel CNOT gates (see Fig.~\ref{fig:chain-12}), and follow the same learning and validation procedure as in the four-qubit case. However, we restrict attention to the non-negative (Markovian) and unconstrained Pauli pseudo-Lindblad models due to the computational difficulty of enforcing complete positivity. Due to the larger system size, the learned generator contains a substantially larger number of Lindblad rates, which allows us to examine the distribution of inferred rates in greater detail. As shown in Fig.~\ref{fig:chain-12}a, the unconstrained model exhibits a broad distribution of rates, including many small negative values as well as a smaller number of more pronounced negative contributions.

We compare the predictions of the two models against experimentally measured expectation values of nonlocal Pauli observables. As shown in Fig.~\ref{fig:chain-12}b, the unconstrained Pauli pseudo-Lindblad model again provides improved agreement with the experimental data compared to the non-negative-rate model. While without a CP-constrained model we cannot conclusively say that this implies the error channel is non-Markovian, the improved predictive power of the unconstrained model argues strongly in this direction.
\section{Discussion}

In this work we have shown that non-Markovianity, as defined by CP-indivisibility, is common for Pauli channels, occurring not just in contrived examples but ubiquitously in random Pauli channels as well as for experimentally relevant error sources. Our theoretical findings are backed up by experimental evidence demonstrating improved accuracy of non-Markovian noise models on IBM Quantum devices. We expect this to have the largest impact in the near-term in quantum error mitigation, and we have generalized two leading candidate protocols to accommodate non-Markovian generators.

We emphasize again that the definition of Markovianity used in this work relates to 
the divisibility of the channel, and therefore depends on whether its generator is of 
GKSL form. It does not imply temporal correlations of error across a quantum circuit. 
Moreover, as we examine the effective error extracted from noisy physical evolution, 
and shaped into a Pauli channel by twirling, our non-Markovianity also does not imply 
temporal correlations within the physical error process during the evolution of the 
noisy quantum gate.

In this setting, it is also important to note that error channels learned in the 
presence of state preparation and measurement errors can only be identified up to 
gauge degrees of freedom~\cite{chen2023learnability,chen2024efficient,chen2025disambiguating}. 
Consequently, this notion of non-Markovianity can be gauge-dependent, in the sense 
that the same physical noise process may appear non-Markovian in one gauge while 
admitting a Markovian representation in another.

Nevertheless, this version of non-Markovianity may complicate aspects of 
fault-tolerant quantum computation, particularly effective noise modeling and 
noise-aware decoding. In particular, recent work has shown that logical qubits can 
exhibit emergent non-Markovian dynamics even when the underlying physical noise is 
Markovian~\cite{kwiatkowski2025constructing,ziyad2025emergent}, and Pauli-Lindblad noise models are increasingly used 
to learn effective logical noise from syndrome 
data~\cite{zheng2026efficient}. The Pauli pseudo-Lindblad framework developed here provides 
a natural tool for capturing such non-Markovian features in these settings. At the 
same time, explicitly accounting for such non-Markovian structure enables more 
accurate and operationally useful noise models, which can inform mitigation and 
characterization approaches beyond strictly Markovian assumptions while remaining 
experimentally scalable.

\textit{Note added.} During the final preparation of this manuscript, we became aware of Ref.~\cite{2026nonmarkovianpauli}, which studies a similar phenomenon.
\acknowledgments
We are grateful to Kristan Temme for early discussions on the foundation, impact, and direction of this work. We also thank Abhinav Kandala for helpful discussions and Ewout van den Berg for helpful discussions and constructive feedback on the manuscript. 

The results presented in this work are the subject of a patent application (Application No.~19/191306), filed on 28 April 2025.

\pagebreak

\appendix
\section{Fitting local Lindbladians}\label{app:learning}
In the experiments we follow the recipe of Ref.~\cite{van2023probabilistic} to fit a local generator with rates $\lambda_k$ to local fidelities $f_k$ of the noise channel, which we review in the following.  Let $\mathcal{P}_{\rm{loc}}$ denote the basis set of 2-local nearest-neighbor Pauli operators. We extract $f_P$ for $P\in\mathcal{P}_{\rm{loc}}$ using standard benchmark techniques such as cycle benchmarking~\cite{erhard2019characterizing}. For each $P$, we prepare one of its eigenstates, apply the unitary of interest, e.g., a layer of CNOT gates, for an even number of times $d$, and measure the expectation value of $P$. From the decay rate of the expectation value, we extract the corresponding fidelity in a way that is robust to state preparation and measurement errors. Note that in this work we focus on fidelities of an even number of CNOT layers to avoid the complexities associated with the gauge-freedom in the Pauli channels~\cite{chen2023learnability,chen2024efficient,chen2025disambiguating}. Let $\vec{f}_{\rm{loc}}$ denote the extracted local fidelities and $\vec{\lambda}_{\rm{loc}}$ denote the Lindblad rates in the model. As shown in Ref.~\cite{van2023probabilistic}, in the ideal case
\begin{equation}\label{eq:designmat}
    M \vec{\lambda}_{\rm{loc}} = \log(\vec{f}_{\rm{loc}})/2, 
\end{equation}
where $M$, the design matrix, is given by
\begin{equation}
    M_{P,Q} = \begin{cases}
        0 & \text{if } [P,Q]=0  \\ 
        1 & \text{otherwise}
    \end{cases},
\end{equation}
for $P,Q\in\mathcal{P}_{\rm{loc}}$. 

We consider three different strategies for finding $\vec{\lambda}_{\rm{loc}}$ (and drop the ``loc" subscript for brevity).

\begin{enumerate}
    \item Constrained least-squares with enforcing positivity on the rates so that
\begin{equation}
\vec{\lambda}^{+} = \underset{\vec{\lambda}\geq0}{{\rm{argmin}}}   \norm{M \vec{\lambda} - \log(\vec{f})/2}^2.
    \end{equation}
    This is the strategy used in Ref.~\cite{van2023probabilistic}
    \item Unconstrained least-square, that is
    \begin{equation}
    \vec{\lambda}^{{\rm{unc}}} = \underset{\vec{\lambda}}{{\rm{argmin}}}  \norm{M \vec{\lambda} - \log(\vec{f})/2}^2.
    \end{equation}
    \item Constrained fit, enforcing complete positivity on the full noise channel, that is
    \begin{equation}
    \vec{p}^{{\rm{cp}}} = \underset{\vec{p}\geq 0, \sum_i p_i =1}{\rm{argmin}} \norm{H \vec{p} - \vec{f}_{\rm{ext}}}^2,
    \end{equation} 
    where $\vec{f}_{\rm{ext}}$ is the vector of all  system fidelities, obtained from the model $\vec{\lambda}^{\rm{unc}}$ and $H$ is the Walsh-Hadamard transform introduced earlier. We then use Eq.~\eqref{eq:lambda-log-fid} to obtain $\vec{\lambda}^{{\rm{cp}}}$, which is guaranteed to generate a CPTP map.  %
    Note that this approach is not computationally efficient as it requires optimization over all $4^n$ Pauli error probabilities for an $n$-qubit system. 
\end{enumerate}

In all experiments, we extract $f_P$ by measuring the corresponding observable after $d=0,2,4,8$ applications of the layer. We use 250 randomizations for twirling, and measure each instance 40 times. The four‑ and twelve‑qubit experiments were performed on  \emph{ibm\_pinguino} using qubits $\{3,4,5,6\}$ and $\{2,3,\dots,13\}$, respectively. 
\section{Random Pauli channels\label{app:random}}

In this Appendix, we present a perturbative derivation of ${\rm{Pr}}(\lambda_{\min})$ and its expectation value for random Pauli channels considered in Sec.~\ref{sec:randompauli}.

Since we consider low-error channels, $r\ll 1$. Therefore, $p_k\ll1$ and hence $f_k$ are close to 1. We can then expand $\log(f_k)$ around 1 to second order in $p_k$ and find 
\begin{align}
    \log(\vec{f}) &= \log(H \vec{p}) \\
    &= H \vec{p} - \vec{1} - \frac{1}{2} (H \vec{p} - \vec{1})\odot (H \vec{p} - \vec{1}) \\
    &= H \vec{p} - \vec{1} - \frac{1}{2} H \vec{p} \odot  H \vec{p} - \frac{1}{2} \vec{1} + H \vec{p} \\
    & = 2 H \vec{p} - \frac{3}{2} \vec{1} - \frac{1}{2} H \vec{p} \odot H \vec{p},
\end{align}
where $\vec{1}$ is a vector with all of its elements equal to 1. Next, we perform the inverse Walsh-Hadamard transform and find 
\begin{align}
    \vec{\lambda} &= \frac{1}{D} H \log(H \vec{p}) \\
    &= 2 \vec{p} - \frac{3}{2} \vec{\delta} - \frac{1}{2D} H ( H \vec{p} \odot H \vec{p}),
\end{align}
where $\vec{\delta}$ is a vector whose zeroth element is 1 and the rest are all 0. We remind the reader that $D=4^n$ is the dimension of an $n$-qubit PTM. To understand the implications of the last term on the right hand side, it is instructive to expand it and find
\begin{align}
    [H(H \vec{p} \odot H \vec{p})]_i &= \sum_j h_{ij} (\sum_k h_{jk} p_k)^2\\
    &= \sum_{j,k,l} h_{ij}h_{jk}h_{jl} p_k p_l 
\end{align}

Examining the elements, we observe that 
\begin{equation}\label{eq:lambda2nd}
\lambda_k = 
    \begin{cases}
        2 p_0 -\frac{3}{2} -\frac{1}{2}\sum_j p_j^2 & \text{if } k=0 \\
        2 p_k - p_0 p_k - \frac{1}{2} \sum_{j\neq 1} p_j p_{g_k(j)} & \text{otherwise,}
    \end{cases}
\end{equation}
where we used the orthogonal property of Hadamard matrices to simplify the expressions. As element-wise multiplication of rows of a Hadamard matrix by another row results in a permutation of the rows, we introduced $g_k$, to denote the permutation when multiplying $H$ by the $k$th row. 

To simplify the analysis, we lift the restriction that $\sum_{k>0}p_k=r$. Instead, we choose $p_k$ for $k>0$ independently at random, and argue that due to the concentration of measure phenomenon this constraint is approximately satisfied for large $n$ for a wide class of  distributions that $p_k$ can be chosen from. Moreover, other quantities involving a sum of many terms, such as the norm, $\sum_k p_k^2$ and $\sum_{kj} p_k p_j$ also concentrate. In particular, we let $p_k = \frac{r}{(D -1)\mu_1} x_k$,  where $x_k$ is chosen at random from a bounded distribution with moments $\mu_\ell$. Since $x_k$ are bounded and independent, using Chernoff inequality we find that
\begin{equation}
    \sum_{k>0} p_k = r + O\left(\frac{\sqrt{\mu_2-\mu_1^2}}{D}\right),
\end{equation}
and 
\begin{equation}
    \sum_{k>0} p_k^2 = \frac{r^2}{D -1} \frac{\mu_2}{\mu_1^2} + O\left(\frac{\sqrt{\mu_4-\mu_2^2}}{D^{3}\mu_1^4 }\right) 
\end{equation}
and  
\begin{equation}
    \sum_{j\neq k,0} p_j p_{g_k(j)} =   \frac{r^2}{(D -1)^2} (D -2) + O\left(\frac{\sqrt{\mu_2^2-\mu_1^4}}{D^{3}\mu_1^4 }\right)
\end{equation}
Therefore, by neglecting fluctuations of $O(\frac{1}{D^2})$ we simplify Eq.~\eqref{eq:lambda2nd} and find 
\begin{equation}\label{eq:lambdasimp}
\lambda_k = 
    \begin{cases}
        -\frac{r}{2}(2+r+\frac{r}{D-1}\frac{\mu_2}{\mu_1^2}) & \text{if } i=0 \\
        (1+r)p_k - \frac{1}{2}  \frac{r^2}{(D -1)^2} (D -2) & \text{otherwise}
    \end{cases},
\end{equation}
We can now see that the negative rates $\lambda_k$ for $k>0$ appear if the first term in the second line of Eq.~\eqref{eq:lambdasimp} is smaller than the second term. 

To find the probability of having a negative rate at large $n$, we find the probability of $\min(p_k)<\frac{r^2}{2(1+r)D}$ for $k>0$. 
Let 
$p_{\min}$ denote $\min(p_k)$  for $k>0$ and define $C_{r,n} = \frac{r^2}{2(1+r)D}$. Therefore, we have 
\begin{align}
   {\rm{Pr}}(p_{\min}<C_{r,n})&= 1 - {\rm{Pr}}(p_{\min}>C_{r,n})\\
    &= 1 - [{\rm{Pr}}(p_k>C_{r,n})]^{D}\\
    &= 1 - [{\rm{Pr}}(\frac{r}{D \mu_1}x_i>C_{r,n})]^{D} \\
    &= 1 - [{\rm{Pr}}(x_i>\frac{\mu_1 r}{2(1+r)})]^{D}
\end{align}

So far, the results are general and we have not chosen a specific distribution for $p_k$. However, we can find the probability of having a negative rate if we are given a specific distribution. 

For example, if $x_k$ are chosen from the uniform distribution over $\mathcal{U}(0,1)$ with $\mu_1 = 1/2$, we have 
\begin{align}\label{eq:prob-min-lambda}
   {\rm{Pr}}(p_{\min}<C_{r,n})
    &= 1 - [{\rm{Pr}}( x_i >\frac{r}{4(1+r)})]^{D} \\
    &= 1 - (1-\frac{r}{4(1+r)})^{D} \\
    &\approx 1-\exp(-\frac{D r}{4(1+r)})
\end{align}

Note that a simple modification to the above calculation allows us to find ${\rm{Pr}}(\lambda_{\min}<y)$, which can be used to find $\expval{\lambda_{\min}}$ as  
\begin{equation}
    \expval{\lambda_{\min}} = \frac{r}{2 D^2-1} \left(-\left(\left(D-4\right) r\right)-4 e^{-D} (r+1)+4\right).
\end{equation}
We compare our finding with numerical simulations and find a good agreement as shown in Fig.~\ref{fig:rand-channel}. 

\section{Twirled Physical noise channels}
\label{appendix:twirled_phys}
In this Appendix, we derive the effective generator of the twirled noise channel for the physically motivated examples discussed in Sec.~\ref{subsec:twirled_phys}.
For each case, we first compute the noisy time-evolution operator and obtain the corresponding Pauli transfer matrix (PTM) of the twirled channel. We then extract the Pauli pseudo-Lindblad noise generator $\mathcal{L}$, and the corresponding parameters $\lambda_k$, by taking the matrix logarithm as in Eqs.~(\ref{eqn:E_from_fk})--(\ref{eq:lambda-log-fid}) (see also Ref.~\cite{malekakhlagh2025efficient}). 

\subsection{$X_{\pi}$ gate with amplitude and phase error}
The time-evolution operator for the noisy Hamiltonian  
\begin{equation}
    H_X = (\Omega+\delta \Omega) X + \epsilon Z, 
\end{equation}
at time $\Omega t = \pi/2$ can be expressed as
\begin{equation}
\begin{split}
U = 
\left(
\begin{array}{cc}
 \cos \left(\frac{\pi  \zeta }{2}\right)-\frac{i \delta_z  \sin
   \left(\frac{\pi  \zeta }{2}\right)}{\zeta } & -\frac{i (\delta_x 
   +1) \sin \left(\frac{\pi  \zeta }{2}\right)}{\zeta } \\
 -\frac{i (\delta_x +1) \sin \left(\frac{\pi  \zeta
   }{2}\right)}{\zeta } & \cos \left(\frac{\pi  \zeta }{2}\right)+\frac{i \delta_z \sin \left(\frac{\pi  \zeta }{2}\right)}{\zeta } 
\end{array}
\right)
\end{split} ,
\end{equation}
where due to the $X$ over-rotation and off-axis $Z$ rotation, there appears a collective normalized rotation frequency $\zeta$ as
\begin{equation}
\zeta \equiv \sqrt{(\delta_x+1)^2+\delta_z^2},
\end{equation}
with $\delta_x \equiv \delta\Omega/\Omega$ and $\delta_z \equiv \epsilon/\Omega$.

We then compute the twirled noise channel $\mathcal{E}=\exp(\mathcal{L})$, the corresponding Pauli fidelities, and the Pauli pseudo-Lindblad generator parameters $\mathcal{L}=\sum_{k} \lambda_k (P_k \bullet P_k - \bullet)$.

The Pauli fidelities are found as
\begin{align}
&f_{I}=1, \\
&f_{X} = \frac{\left(\delta _x+1\right){}^2+\cos (\pi\zeta ) \delta _z^2}{\zeta^2}, \\
& f_{Y}=-\cos (\pi  \zeta ) \\
&f_{Z} = -\frac{\cos (\pi  \zeta ) \left(\delta _x+1\right){}^2+\delta _z^2}{\zeta
   ^2}
\end{align}
and the corresponding Pauli pseudo-Lindblad parameters read
\begin{align}
\begin{split}
\lambda_X & = +\frac{1}{4}\log\left(\left(\delta _x+1\right){}^2+\cos (\pi  \zeta ) \delta_z^2\right)\\
&-\frac{1}{4}\log (-\cos (\pi  \zeta ))\\
& -\frac{1}{4}\log \left(-\cos (\pi  \zeta
   ) \left(\delta _x+1\right){}^2-\delta _z^2\right) \;,
\end{split}
\end{align}
\begin{align}
\begin{split}
\lambda_Y & = \frac{1}{4} \log (-\cos (\pi  \zeta ))\\
&-\frac{1}{4}\log \left(-\cos (\pi \zeta) \left(\delta _x+1\right){}^2-\delta _z^2\right)\\
&-\frac{1}{4}\log\left(\left(\delta _x+1\right){}^2+\cos (\pi  \zeta ) \delta_z^2\right)\\
&+\frac{1}{2} \log \left(\left(\delta _x+1\right){}^2+\delta_z^2\right) \;,
\end{split}
\end{align}
\begin{align}
\begin{split}
\lambda_Z & = -\frac{1}{4}\log (-\cos (\pi  \zeta ))\\
& + \frac{1}{4}\log \left(-\cos (\pi \zeta) \left(\delta _x+1\right){}^2-\delta _z^2\right) \\ 
&-\frac{1}{4}\log\left(\left(\delta _x+1\right){}^2+\cos (\pi \zeta ) \delta_z^2\right) \;.
\end{split}
\end{align}
The behavior of the pseudo generator parameters, particularly where they change sign, is shown in Fig.~\ref{fig:x-gate} of the main text.

\subsection{$ZZ_{\pi/2}$ gate with $T_1$ damping}
Consider a $ZZ_{\pi/2}$ gate with continuous-time $T_1$ dissipator on both qubits as
\begin{equation}
    \partial_t\rho = - iJ [Z_1 Z_2,\rho] + \sum_{j=1,2} \kappa (\sigma^-_j \rho  \sigma^+_j - \frac{1}{2} \{\sigma^+_j\sigma^-_j, \rho\}).
\end{equation}
Here, we present the Pauli fidelities and the corresponding Pauli pseudo-Lindblad generator parameters as a function of the relative noise strength $\delta_{\kappa}\equiv \kappa/4J$. 

The Pauli fidelities of the twirled noise channel can be grouped as 
\begin{align}
f_{II} & = 1, \\
f_{IX} & = f_{IY} = f_{XI} = f_{YI} = \frac{e^{-\frac{3 \pi  \delta _{\kappa }}{2}} \left(e^{\pi  \delta _{\kappa }}+1\right)}{2 \left(\delta _{\kappa }^2+1\right)},\\
\begin{split}
f_{IZ}&=f_{XX}=f_{XY}=\\
f_{YX}&=f_{YY}=f_{ZI}= e^{-\pi \delta_{\kappa}} ,
\end{split} \\
\begin{split}
f_{XZ}&=f_{YZ}=f_{ZX}=f_{ZY}\\
&=\frac{e^{-\frac{3 \pi  \delta _{\kappa }}{2}} \left(e^{\pi  \delta _{\kappa }}+1\right) \left(2 \delta _{\kappa }^2+1\right)}{2
   \left(\delta _{\kappa }^2+1\right)} \;,
\end{split}\\
f_{ZZ} & = e^{- 2\pi \delta_{\kappa}} \;.
\end{align}
The corresponding $\lambda_k$ parameters are obtained as 
\begin{align}
\begin{split}
\lambda_{IX} & = \lambda_{IY} = \lambda_{XI} = \lambda_{YI} \\
&= \frac{1}{8} \left(\pi  \delta _{\kappa }-\log \left(2 \delta _{\kappa }^2+1\right)\right) \\
& = \frac{\pi  \delta _{\kappa }}{8} -\frac{\delta _{\kappa }^2}{4} + \frac{\delta _{\kappa }^4}{4} + O(\delta_{\kappa}^5) \;,
\end{split}
\end{align}
\begin{align}
\begin{split}
&\lambda_{IZ}=\lambda_{XX}=\lambda_{XY}=\\
&\lambda_{YX}=\lambda_{YY}=\lambda_{ZI}= 0 \;,
\end{split}
\end{align}
\begin{align}
\begin{split}
\lambda_{XZ}&=\lambda_{YZ}=\lambda_{ZX}=\lambda_{ZY} \\
& = \frac{1}{8} \left(\pi  \delta _{\kappa }+\log \left(2 \delta _{\kappa }^2+1\right)\right) \\
& = \frac{\pi  \delta _{\kappa }}{8} +\frac{\delta _{\kappa }^2}{4} - \frac{\delta _{\kappa }^4}{4} + O(\delta_{\kappa}^5) \;,
\end{split}
\end{align}
\begin{align}
\begin{split}
\lambda_{ZZ} & = \frac{1}{4} \Big[\log (4) + \pi  \delta _{\kappa }-2 \log \left(e^{\pi  \delta _{\kappa }}+1\right)\\
&+ 2 \log \left(\delta _{\kappa }^2+1\right)
-\log\left(2 \delta _{\kappa }^2+1\right)\Big] \\
& = -\frac{1}{16} \pi ^2 \delta_{\kappa }^2 + \frac{1}{384} \left(96+\pi ^4\right) \delta _{\kappa }^4 +O(\delta_{\kappa}^5) \;.
\end{split}
\label{Eq:ZZgateWithT1-lambda_ZZ}
\end{align}
where the last lines show the series expansion in $\delta_{\kappa}$ up to $O(\delta_{\kappa}^5)$. 

The expressions here are plotted in Fig.~\ref{fig:rzz-gate} of the main text. In particular, note that $\lambda_{ZZ}$ is negative even for very weak noise as found in Eq.~(\ref{Eq:ZZgateWithT1-lambda_ZZ}).   
\bibliographystyle{apsrev4-2}
\bibliography{nonmarkov}
\end{document}